\documentclass[aps,prl,twocolumn,groupedaddress,showpacs]{revtex4}

\usepackage{bm}
\usepackage{graphicx}

\begin{document}

\title{Band Structure and Fermi Surface of an Extremely Overdoped Iron-Based Superconductor KFe$_2$As$_2$}
\author{T. Sato,$^{1,2}$ K. Nakayama,$^1$ Y. Sekiba,$^1$ P. Richard,$^3$ Y.-M. Xu,$^4$ S. Souma,$^3$ T. Takahashi,$^{1,3}$ G. F. Chen,$^5$ J. L. Luo,$^5$ N. L. Wang,$^5$ and H. Ding$^5$}
\affiliation{$^1$Department of Physics, Tohoku University, Sendai 980-8578, Japan}
\affiliation{$^2$TRIP, Japan Science and Technology Agency (JST), Kawaguchi 332-0012, Japan}
\affiliation{$^3$WPI Research Center, Advanced Institute for Materials Research, 
Tohoku University, Sendai 980-8577, Japan}
\affiliation{$^4$Department of Physics, Boston College, Chestnut Hill, MA 02467, USA}
\affiliation{$^5$Beijing National Laboratory for Condensed Matter Physics, and Institute of Physics, Chinese Academy of Sciences, Beijing 100190, China}

\date{\today}

\begin{abstract}	
We have performed high-resolution angle-resolved photoemission spectroscopy on heavily overdoped KFe$_2$As$_2$ (transition temperature $T_{\rm c}$ = 3 K).  We observed several renormalized bands near the Fermi level with a renormalization factor of 2-4.  While the Fermi surface (FS) around the Brillouin-zone center is qualitatively similar to that of optimally-doped Ba$_{1-x}$K$_x$Fe$_2$As$_2$ ($x$ = 0.4; $T_{\rm c}$ = 37 K), the FS topology around the zone corner (M point) is markedly different: the two electron FS pockets are completely absent due to excess of hole doping.  This result indicates that the electronic states around the M point play an important role in the high-$T_{\rm c}$ superconductivity of Ba$_{1-x}$K$_x$Fe$_2$As$_2$ and suggests that the interband scattering via the antiferromagnetic wave vector essentially controls the $T_{\rm c}$ value in the overdoped region.
\end{abstract}

\maketitle

      The discovery of superconductivity at 26 K \cite{Kamihara} (43 K under high pressure \cite{Takahashi}) in LaFeAsO$_{1-x}$F$_x$, has triggered intensive researches on the high-temperature ($T_{\rm c}$) superconductivity of iron (Fe) pnictides.  The $T_{\rm c}$ value has already exceeded 55 K by replacing La atom with other rare-earth atoms or by introducing oxygen vacancies\cite{AIST, RenCPL}, opening a new avenue for high-$T_{\rm c}$ material research beside cuprates.  Remarkable aspects of the FeAs-based superconductors are (i) electrons in the Fe orbitals, generally believed to be the foe, indeed play an essential role in superconductivity\cite{Kamihara, SatoJPSJ, HongEPL}, (ii) non-doped parent compounds commonly exhibit a collinear antiferomagnetic (AF) spin density wave (SDW)\cite{WangPRL, DaiNature}, and (iii) the superconductivity emerges either by the hole or electron doping into the parent compounds\cite{Kamihara, Rotter}.  To elucidate the mechanism of high-$T_{\rm c}$ superconductivity in terms of the electronic structure, angle-resolved photoemission spectroscopy (ARPES) has been performed on both hole- and electron-doped compounds in the optimally- and non(under)-doped region \cite{HongEPL, Kondo, AdamNd, FengSr, AdamBK, ZhouBK, ShenP, Borisenko, Hasan} and it clarified key features on the band structure, the FS topology, and the superconducting gap.  On the other hand, little is known about the electronic states in the overdoped region.  As demonstrated by electrical resistivity measurements, the $T_{\rm c}$ value of the hole-doped Ba$_{1-x}$K$_x$Fe$_2$As$_2$ monotonically decreases from the optimally-doped region ($T_{\rm c}$ = 37 K) upon hole doping but does not completely disappear even at the highest doping level ($x$ = 1.0; $T_{\rm c}$ $\sim$3 K)\cite{Rotter2, ChenH}, unlike the overdoped cuprates.  The resistivity does not show SDW-related anomalies in the overdoped region\cite{ChenH}.  Clarifying the microscopic origin of this $T_{\rm c}$ reduction would be a key to find an essential ingredient to achieve high-$T_{\rm c}$ values in the iron-based superconductors.  It is thus of particular importance to gain insight into the band structure and the FS by performing ARPES measurements on overdoped samples and directly compare the electronic states with the optimally-doped ones for a comprehensive understanding of the high-$T_{\rm c}$ mechanism.
	  
In this Letter, we report high-resolution ARPES results on KFe$_2$As$_2$ ($T_{\rm c}$ = 3 K), the overdoped limit of $A$$_{1-x}$K$_x$Fe$_2$As$_2$ ($A$ = Alkali earth metal; $x$ = 1).  We have determined the band structure near $E_{\rm F}$ and the FS topology, and compared with the results of optimally-doped Ba$_{0.6}$K$_{0.4}$Fe$_2$As$_2$ ($T_{\rm c}$ = 37 K).  We demonstrate that, unlike Ba$_{0.6}$K$_{0.4}$Fe$_2$As$_2$, the nesting condition via the AF wave vector is not satisfied in KFe$_2$As$_2$.  We discuss the implications of the present results in terms of electron correlations and interband scattering.

High-quality single crystals of KFe$_2$As$_2$ used in this study were grown by the flux method\cite{ChenGF}. High-resolution ARPES measurements were performed using a VG-SCIENTA SES2002 spectrometer with a high-flux discharge lamp and a toroidal grating monochromator.  We used the He I$\alpha$ resonance line ($h$$\nu$ = 21.218 eV) to excite photoelectrons.  The energy and angular (momentum) resolutions were set at 4-10 meV and 0.2$^\circ$ (0.007$\AA$$^{-1}$), respectively.  Clean surfaces for ARPES measurements were obtained by {\it in-situ} cleaving of crystals in a working vacuum better than 5$\times$10$^{-11}$ Torr.  The Fermi level ($E_{\rm F}$) of the samples was referenced to that of a gold film evaporated onto the sample substrate.  Mirror-like sample surfaces were found to be stable without obvious degradation for the measurement period of 3 days.

\begin{figure}
\includegraphics[width=3.4in]{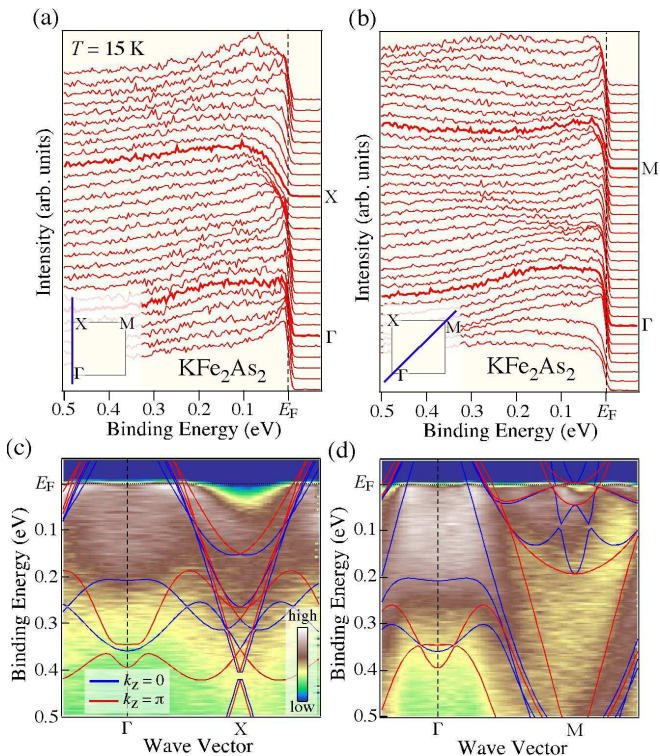}
\caption{(Color online) ARPES spectra near $E_{\rm F}$ of KFe$_2$As$_2$ ($T_{\rm c}$ = 3 K) measured at 15 K with the He I$\alpha$ line along two high symmetry lines (a) $\Gamma$X and (b) $\Gamma$M.  (c) and (d), ARPES intensity plots as a function of wave vector and binding energy along the $\Gamma$X and $\Gamma$M lines, respectively, together with the band dispersion from the LDA calculations for $k_{\rm z}$ = 0 and $\pi$ (blue and red curves, respectively).  Calculated bands for BaFe$_2$As$_2$ \cite{ZFang} are shifted upward by 200 meV and then renormalized by a factor of 2.}
\end{figure}

  Figure 1 shows ARPES spectra of KFe$_2$As$_2$ measured at 15 K with the He I$\alpha$ line along two high-symmetry lines (a) $\Gamma$X and (b) $\Gamma$M.  In the $\Gamma$X cut (Fig. 1(a)), we find a band showing a holelike dispersion centered at the $\Gamma$ point which crosses $E_{\rm F}$ midway between the $\Gamma$ and X points.  This band has a bottom of dispersion around 0.1 eV at the X point.  In the $\Gamma$M cut (Fig. 1(b)), we observe a few less-dispersive bands crossing $E_{\rm F}$ which always stay within 0.1 eV with respect to $E_{\rm F}$.  We also observe a broad feature rapidly dispersing toward higher binding energy on approaching the M point.  To see more clearly the dispersive bands, we have mapped the ARPES intensity as a function of wave vector and binding energy and show the results in Figs. 1(c) and (d) for the $\Gamma$X and $\Gamma$M directions, respectively.  We also plot the band structure calculated within the local-density approximation (LDA) at $k_{\rm z}$ = 0 and $\pi$ (blue and red curves, respectively)\cite{ZFang}.  The calculated bands for BaFe$_2$As$_2$ are shifted upward by 0.2 eV to account for the chemical-potential shift due to hole doping, and then divided by a renormalization factor of 2.  As clearly seen in Figs. 1(c) and (d), the experimentally determined band structure shows an overall agreement with the renormalized LDA calculations.  Especially, a highly-dispersive band in Fig. 1(d) whose energy exceeds 0.5 eV at midway between the $\Gamma$ and M points is well reproduced by the renormalized calculations. 	We also find some differences in the energy position of bands between the experiment and the calculations, {\it e.g.}, the higher-energy bands at the $\Gamma$ point (0.1-0.2 eV in the experiment {\it v.s.} 0.2-0.4 eV in the calculations), suggesting that some bands are more strongly renormalized with a renormalization factor of $\sim$4.  This implies a possible orbital/momentum dependence of the electron correlation effect.
  
\begin{figure}
\includegraphics[width=3.4in]{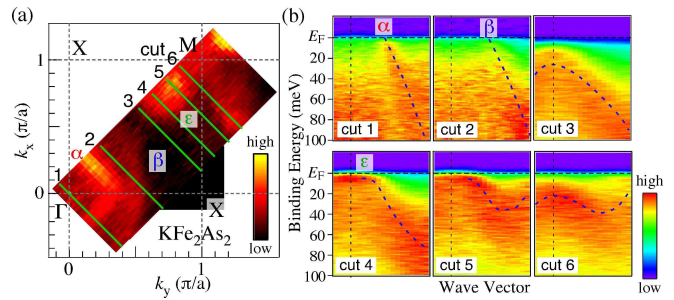}
\caption{(Color online) (a) ARPES-intensity plot at $E_{\rm F}$ of KFe$_2$As$_2$ as a function of the two-dimensional wave vector measured at 15 K.  The intensity at $E_{\rm F}$ is obtained by integrating the spectra within $\pm$10 meV with respect to $E_{\rm F}$.  (b) Representative ARPES-intensity plots in the vicinity of $E_{\rm F}$ as a function of binding energy and wave vector measured along cuts 1-6 indicated by green lines in (a).  Blue dashed lines are guide for eyes to trace the band dispersion.  Black dashed line indicates the momentum on the $\Gamma$M line.}
\end{figure}

Figure 2(a) shows the ARPES-intensity plot at $E_{\rm F}$ as a function of the two-dimensional wave vector measured at 15 K.  Bright areas correspond to the experimentally-determined FS.  We identify two holelike FSs centered at the $\Gamma$ point, where the inner one displays a stronger intensity.  These FSs are also observed in Ba$_{0.6}$K$_{0.4}$Fe$_2$As$_2$ \cite{HongEPL} (called K0.4 sample) and therefore ascribed to the $\alpha$ (inner) and $\beta$ (outer) hole pockets.  We also find small bright spots slightly away from the M point, called here the $\epsilon$ pockets.  To clarify the topology of the $\epsilon$ FS, we plot in Fig. 2(b) the ARPES intensity in the vicinity of $E_{\rm F}$ as a function of the wave vector and binding energy for representative cuts (1-6).  As seen in the cuts 3-6, the band to produce the $\epsilon$ FS is less dispersive than the $\alpha$ and $\beta$ bands near the $\Gamma$ point (cuts 1 and 2, respectively) and shows a holelike dispersion with respect to the $\Gamma$M line (black dashed line).  This $\epsilon$ band crosses $E_{\rm F}$ in cuts 4 and 5, producing a small hole pocket.  No evidence for the two electron pockets ($\gamma$ and $\delta$ FSs) \cite{HongEPL, ZFang, Singh, JDai, FMa, Nekrasov} has been found at the M point, likely due to the excess of hole doping which lifts the band bottom of the electron pockets above $E_{\rm F}$.

\begin{figure}
\includegraphics[width=3.0in]{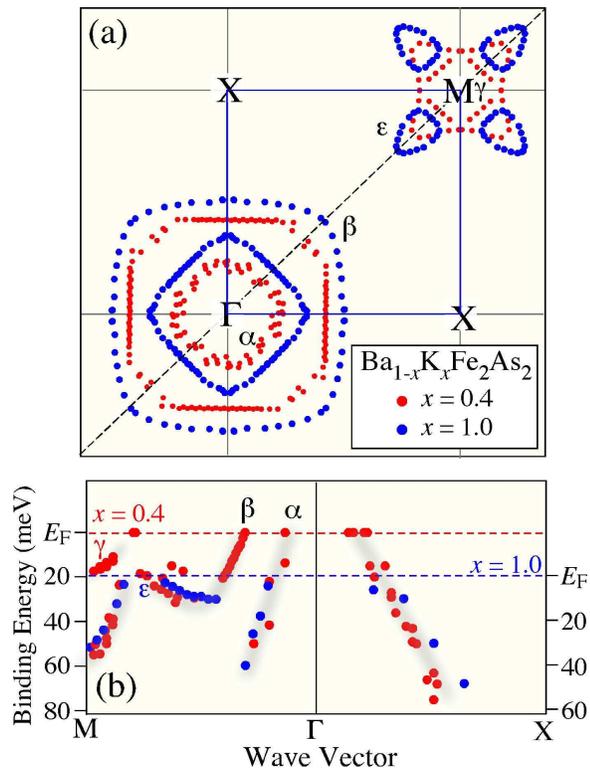}
\caption{(Color online) (a) Comparison of experimentally determined $k_{\rm F}$ points between overdoped KFe$_2$As$_2$ ($T_{\rm c}$ = 3 K) and optimally-doped Ba$_{0.6}$K$_{0.4}$Fe$_2$As$_2$ ($T_{\rm c}$ = 37 K) \cite{HongEPL} (blue and red circles, respectively). The $k_{\rm F}$ points are symmetrized by assuming a four-fold symmetry with respect to the $\Gamma$ and M points.  (b) Experimental band dispersion in the vicinity of $E_{\rm F}$ for two high-symmetry lines determined by tracing the peak position of the ARPES spectra.  The chemical potential of the K1.0 sample is shifted downward with respect to that of the K0.4 sample.}
\end{figure}

\begin{figure}
\includegraphics[width=3.2in]{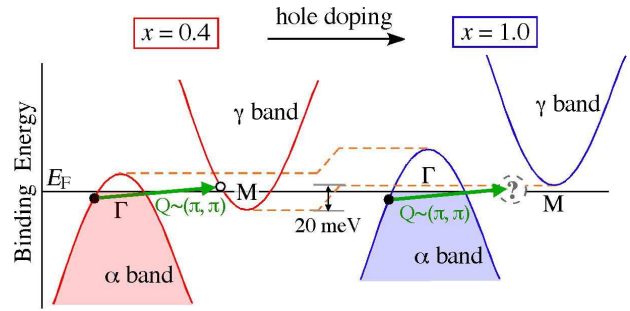}
\caption{(Color online) Schematic view of the interband scattering by the AF wave vector $Q_{\rm AF}$ between the hole and electron bands (the $\alpha$ and $\gamma$ bands) centered at $\Gamma$ and M points, respectively.  The interband scattering is markedly suppressed in the overdoped region due to the absence of the electron FSs at M.}
\end{figure}

Figure 3(a) displays a direct comparison of the momentum location of $k_{\rm F}$ points between the K0.4 and K1.0 samples.  As clearly visible, the volume of the $\alpha$ and $\beta$ FSs expands due to hole doping.  Indeed, the estimated volumes of these two FSs with respect to the first BZ for $x$ = 1.0 are 7 and 22$\%$, respectively, much larger than that for $x$ = 0.4 (4 and 18 $\%$, respectively \cite{HongEPL}).  By taking into account the volume of the $\epsilon$ FS (2.5 $\%$) and the predicted degenerated $\beta$ band \cite{ZFang} (22 $\%$), the total hole concentration is estimated to be 1.05.  This value is remarkably consistent with the doping level of $x$ = 1.0, indicating the bulk nature of this measurement.  We emphasize here that the FS topology at the M point is drastically different between the two doping levels since the small $\gamma$ and $\delta$ electron pockets observed at $x$ = 0.4 are completely absent at $x$ = 1.0.  This marked difference is reasonably explained by a simple energy shift due to hole doping.  As shown in Fig. 3(b), the energy position of the bands in the vicinity of $E_{\rm F}$ for $x$ = 0.4 and 1.0 quantitatively matches each other when we shift down the bands for $x$ = 1.0 by 20-25 meV, demonstrating the applicability of a rigid-band model.  This chemical potential shift is much smaller than the expected value from the band calculations ($\sim$120 meV), possibly due to the strong mass renormalization of the near-$E_{\rm F}$ bands as seen in Fig. 1.  Since the bottom of the electronlike $\gamma$ band for $x$ = 0.4 is 18 meV below $E_{\rm F}$, the $\gamma$ band for $x$ = 1.0 completely disappears in the unoccupied region by the observed chemical potential shift of 20-25 meV.

Now we discuss the present ARPES results and their implication on the pairing mechanism.  As seen in the schematic band structure in Fig. 4(a), the $\alpha$ and $\gamma$ FSs of the optimally-doped K0.4 sample are well connected by the AF wave vector $Q_{\rm AF}$ = ($\pi$, $\pi$) in the unreconstructed zone scheme\cite{HongEPL}.  Therefore, the enhanced inter-band scattering via the $Q_{\rm AF}$ vector would promote the kinetic process of pair scattering between these two FSs by low-energy fluctuations, leading to an increase of the pairing amplitude\cite{Mazin, Kuroki,FWang}.  This concept is supported by the observation of a superconducting-gap magnitude that is twice larger for the $\alpha$ and $\gamma$ FSs (12 meV) as compared to the $\beta$ FS (6 meV)\cite{HongEPL}.  On the other hand, in the overdoped region ($x$ = 1.0), this interband scattering would diminish since the $\gamma$ band is in the unoccupied side and the size of the $\alpha$ band increases significantly.  This suggests that the interband scattering between the $\alpha$ and $\gamma$ bands is an essential ingredient to achieve high-$T_{\rm c}$ values in iron-based superconductors.  We further speculate that the $\beta$ FS may not contribute deeply to the pairing mechanism in the hole-doped region since its topology and shape are quite similar between optimally- and over-doped samples whereas the $T_{\rm c}$ values are dramatically different.  The opening of a small superconducting gap of the $\beta$ band of the K0.4 sample \cite{HongEPL} might be due to a proximity effect.

In conclusion, we reported ARPES results on KFe$_2$As$_2$ ($T_{\rm c}$ = 3 K) and determined the band dispersion near $E_{\rm F}$ and the FS topology.  The experimentally determined FS consists of three types of hole pockets, two centered at the $\Gamma$ point ($\alpha$ and $\beta$), and one around the M point ($\epsilon$).  The small $\gamma$ electron pocket seen in the optimally-doped sample is completely absent, resulting in the suppression of the $Q_{\rm AF}$ = ($\pi$, $\pi$) interband scattering which is likely responsible for the reduction of $T_{\rm c}$ values in the overdoped region.

\begin{acknowledgments}
We thank X. Dai, Z. Fang, and Z. Wang for providing their LDA results and valuable discussions. This work was supported by grants from JSPS, TRIP-JST, CREST-JST, MEXT of Japan, the Chinese Academy of Sciences, NSF, Ministry of Science and Technology of China, and NSF of US.
\end{acknowledgments}

\end{document}